\documentclass[preprints,article,accept,oneauthor,pdftex]{Definitions/mdpi} 

\firstpage{1} 
\pubvolume{xx}
\issuenum{1}
\articlenumber{5}
\pubyear{2020}
\copyrightyear{2020}
\history{Received: 26 May 2020; Accepted: 6 July 2020; Published: 10
  July 2020}

\newcommand{\del}{\partial}
\newcommand{\alphaIR}{\alpha_\textsc{ir}}
\newcommand{\alphaWW}{\alpha_\textsc{ww}}
\newcommand{\alphaNG}{\alpha_\textsc{ng}}
\newcommand{\hsp}[1]{\hspace*{#1 mm}}
\newcommand{\Nbar}{\overset{\rule[-0.5mm]{2.5mm}{0.2mm}}{N}}
\def\hhref#1{\href{http://arxiv.org/abs/#1}{arXiv:#1}} 

\usepackage{amssymb,bm}
\usepackage{multicol}

\Title{Genuine Dilatons in Gauge Theories} 

\Author{R.~J.~Crewther 
}     

\AuthorNames{R.~J.~Crewther}   

\address[1]{CSSM, Department of Physics, University of Adelaide,
  Adelaide, SA 5005, Australia; rodney.crewther@adelaide.edu.au}    

\abstract{A genuine dilaton $\sigma$ allows scales to exist
  even in the limit of \emph{exact} conformal invariance. In gauge
  theories, these may occur at an infrared fixed point (IRFP) $\alpha_\textsc{ir}$
  through dimensional transmutation. These large scales at $\alpha_\textsc{ir}$
  can be separated from small scales produced by $\theta^\mu_\mu$, \mbox{the
  trace} of the energy-momentum tensor. For quantum chromodynamics
  (QCD), the conformal limit can be combined with chiral 
  $SU(3) \times SU(3)$ symmetry to produce chiral-scale perturbation
  theory $\chi$PT$_\sigma$, with $f_0(500)$ as the dilaton. The
  technicolor (TC) analogue of this is crawling TC: at low energies,
  the gauge coupling $\alpha$ goes directly to (but does not
  walk past) $\alpha_\textsc{ir}$, and the massless dilaton at $\alpha_\textsc{ir}$ 
  corresponds to a light Higgs boson at $\alpha \lesssim \alpha_\textsc{ir}$.
  It is suggested that the $W^\pm$ and $Z^0$ bosons set the scale of
  the Higgs boson mass. Unlike crawling TC, in walking TC,
  $\theta^\mu_\mu$ produces \emph{all} scales, large and small, so it
  is hard to argue that its ``dilatonic'' candidate for the Higgs
  boson is not heavy. 
}  


\begin{document}



\section{Introduction}\label{intro}

It is surprising how far the notion of a ``dilaton'' has strayed from
the original version of 1968--1970. Now it can be any of the following:
\renewcommand{\labelenumi}{\Roman{enumi}.}  
\begin{enumerate}[leftmargin=*,labelsep=4.9mm]
\item a scalar Nambu–Goldstone (NG) boson for \emph{exact} conformal
          invariance of a Hamiltonian $\cal H$ which has a
          \emph{scale-dependent} ground state $|\text{vac}\rangle$
          and hence scale-dependent amplitudes in the limit of
          scale invariance; or 
\item a scalar component of the gravitational field; or
\item a scalar particle in a theory where conformal invariance 
          is permitted only in the Wigner–Weyl (WW) mode 
          (scale-invariant amplitudes). In~terms of a Hamiltonian 
          $\cal H$, scale-dependent 
          effects such as fermion condensation exist \emph{only} in
          the presence of a term $\delta{\cal H}$ which breaks scale
          invariance explicitly in ${\cal H} = {\cal H}_0+\delta{\cal H}$.      
          Both ${\cal H}_0$ and its ground state $|\text{vac}\rangle_0$ are
         \mbox{ conformal invariant}.
 \end{enumerate}
\renewcommand{\labelenumi}{\arabic{enumi}.}  

Evidently, I and III contradict each other and may have little to do
with II. Typical of III are (a)~deformed conformal Lagrangians and
(b) walking TC, which have been promoted as ways of
explaining why the Higgs boson is so light. I observe that these
theories are very unlikely to achieve this because they apparently
involve just one scale, set by $\theta^\mu_\mu$. Two scales are
needed, as~in crawling TC~\cite{crawl}, which is a type-I~theory.

Most papers on ``dilatons'' consider type-II or type-III. The~problem
is with assertions that type-III theories can be matched to 
type-I effective Lagrangians. So I begin with a quick summary of the
fundamental type-I theory in its original setting, strong interactions
(Section \ref{hadronic}). 
 
Dilatonic versions of gauge theories are considered in
Section~\ref{gauge}. For~quantum chromo\-dynamics (QCD), only a type-I theory
is possible, chiral-scale perturbation theory $\chi$PT$_\sigma$
\cite{ct1,ct2,ct3}, where the dilaton at the IRFP corresponds to the
\emph{light} resonance $f_0(500)$. The~TC analogue of this is crawling
TC, \mbox{where the} Higgs boson is the type-I TC analogue of
$f_0$. The~scale-dependent IRFP lies \emph{outside} the conformal
window~\cite{debbio14}.   

Section~\ref{competing} compares crawling TC with type-III theories
for the Higgs boson. The~type-III concept is due to Gildener and
Weinberg~\cite{gild76}. They called their spin-$0^+$ particle a
``scalon''---a good name---but that morphed into the term ``dilaton''
in ``dilatonic'' walking TC~\cite{bardeen86, yam86, hold87, hold88,
 appel10, yam11, yam14, golt16, golt18} and deformed conformal potential 
theory~\cite{meiss07,chang07,foot07,gold08,vecc10,bell13,bell14,cora13}. 
I will reserve the term ``genuine dilaton'' for type-I~dilatons.

The key observation is that in type-III theories, it is hard to
distinguish small scales from large scales because they are all
generated by the trace $\theta^\mu_\mu$.  For~example, in~walking TC,
the sill of the conformal window produces the fermion condensate
$\langle\overline{\psi}\psi\rangle_\text{vac}$.  
That indicates a \emph{large} mass
\begin{equation}
m_{h^{}_{_\text{III}}}
= \bigl\langle h^{}_{{}_\text{III}} 
    \bigl|\theta^\mu_\mu \bigr| h^{}_{{}_\text{III}} \bigr\rangle 
\sim \ \text{a few TeV}
\end{equation}
for the would-be Higgs boson $h^{}_{{}_\text{III}}$, no matter how slowly
$\alpha$ walks. A~similar conclusion was drawn in early work on
walking TC~\cite{bardeen86,hold87,hold88}. It cannot be undone
by assuming~\cite{gold08} an equivalence to a type-I dilaton Lagrangian
below the~sill.

By contrast, in~a type-I theory, all large scales arise from the scale
dependence of amplitudes in the exact conformal limit: that 
is what is meant by the NG mode for conformal invariance.  In~crawling TC, 
the fermion condensate $\langle\overline{\psi}\psi\rangle_\text{vac}$
sets the scale \emph{at} the IRFP $\alphaIR$. For~values of $\alpha$
just below $\alphaIR$,  $\theta^\mu_\mu$ appears as a small  
perturbation which produces small scales, such as the mass
acquired by the type-I dilaton: a light Higgs~boson. 
 
Why has the concept that there can be scale dependence in the
conformal limit, which seemed so simple in 1968--70, been so
systematically overlooked since then?  In particular, why 
must all IRFP's be in WW mode? An IRFP cannot appear outside 
the conformal window, by~definition? I offer possible reasons for
these points of view in Section~\ref{scales}, with~a separate
Section~\ref{irfp} specifically for~IRFP's. 

Possible tests of these proposals are considered in
Section~\ref{renorm}. A~light scalar boson has been observed in
lattice data for  $SU(3)$ gauge theory with $N_f = 8$ triplet
fermions~\cite{aoki14,aoki17,appel16,appel19}  
and two sextet fermions~\mbox{\cite{fodor14,fodor18}}. In~each case,
this is being interpreted as a type-III dilaton for walking TC, but~it
is more likely to be a genuine dilaton, and~hence evidence for an IRFP
just outside the \mbox{conformal window}. 

\section{Hadronic~Physics}\label{hadronic}

The idea that scale and conformal invariance may be spontaneously
(i.e.,\ \emph{not} explicitly) broken dates from 1962 (footnote 38
of~\cite{mgm62}). An~analogy was drawn with the partial conservation of
the axial-vector currents $\vec{\cal F}_{\mu 5}$ for chiral $SU(3)_L
\times SU(3)_R$ symmetry, where $\pi, K, \bar{K}, \eta$ pole dominance
of amplitudes of the divergences $\del^\mu\vec{\cal F}_{\mu 5}$ yields
soft-meson relations such as the Goldberger–Treiman relation for the
pion-nucleon coupling constant $g_{\pi NN}$. Similarly, a~spin-$0^+$
particle $\sigma$ tied to the trace $\theta^\mu_\mu$ of the
energy-momentum tensor $\theta_{\mu\nu}$ couples universally to
particle mass. For~a nucleon $N$ with mass $M_N$, the~$\sigma$-nucleon
coupling constant $g_{\sigma  NN}$ is given  by
\begin{equation}\label{scalarGT}
f_\sigma g_{\sigma NN} \simeq  M_N \,,
\end{equation}
where $f_\sigma$ is the scalar analogue of the pion decay constant $f_\pi$:
\begin{equation}\label{decay}
\langle{\sigma}|\theta_{\mu\nu}|\text{vac}\rangle = 
(f_\sigma/3)(q_\mu q_\nu - g_{\mu\nu}q^2) \,.
\end{equation}
The currents for scale and conformal transformations can be written
in terms of $\theta_{\mu\nu}$ (improved~\cite{ccj70} if spin-$0$ fields
are present) as follows:
\begin{equation}
{\cal D}_\nu(x) = x^\mu \theta_{\mu\nu}(x) 
 \quad\mbox{and}\quad
{\cal K}_{\mu\nu} = (2x_\mu x_\lambda - x^2 g_{\mu\lambda})
                             \theta^\lambda_\nu \,.
\end{equation}
As for chiral $SU(3)_L \times SU(3)_R$ currents, explicit breaking of 
the symmetry is measured by \mbox{current divergences}:
\begin{equation}
\del^\nu {\cal D}_\nu = \theta^\lambda_\lambda
 \quad\mbox{and}\quad
\del^\nu{\cal K}_{\mu\nu} = 2x_\mu \theta^\lambda_\lambda \,.
\end{equation}
Therefore, scale and conformal invariance correspond to the limit
\begin{equation}
\theta^\lambda_\lambda \to 0.
\end{equation}
The question~\cite{nambu68,mack68,mack69,mgm69} is, does the vacuum
state respect the symmetry, or~break it?  Are scale and conformal invariance 
realized in the WW mode or the ``spontaneous'' NG mode?

A comparison with the chiral $SU(3)_L \times SU(3)_R$ group in
hadronic physics is instructive~\cite{mgm69}. In~that case, both modes
occur. The~subgroup $SU(3)_{L+R}$ associated with vector currents 
$\vec{\cal F}_\mu$ has \mbox{a symmetry} limit in the WW mode---its generators
$\vec{F}$ annihilate the vacuum state:
\begin{equation}
 \vec{F}\,|\text{vac}\rangle \to 0 \ , \quad 
 \vec{F} = \int\!d^3x\,\vec{\cal F}_0  \,.
\end{equation}
The symmetry is manifest: its representations can be seen in 
the particle spectrum. The~rest of the group, represented by cosets
$SU(3)_L \times SU(3)_R\bigl/SU(3)_{L+R}$ and generated by axial
charges $\vec{F}_5$, has its symmetry realized in the NG mode:
\begin{equation}\label{axial}
  \vec{F}_5\,|\text{vac}\rangle \not\to 0 \ , \quad 
 \vec{F}_5 = \int\!d^3x\,\vec{\cal F}_{05}  \,.
\end{equation}
As a result, the~axial part of the symmetry is hidden,
$|\text{vac}\rangle$  becomes a member of a degenerate set of
physically equivalent vacua
\begin{equation}\label{chiral_deg}
|\text{vac}\rangle_{\vec{\alpha}} = \exp\{i\vec{\alpha}\!\cdot\!\vec{F}_5\}
    |\text{vac}\rangle_{\vec{\alpha} = 0}  \,,
\end{equation}
and there is a massless NG boson for each independent direction in
$\vec{\alpha}$ space: for $SU(3)_L \times SU(3)_R$, eight $0^-$ NG
bosons $\pi, K, \bar{K}, \eta$. A~unique vacuum state can be picked
out by perturbing the Hamiltonian with a term which breaks the axial
part of the symmetry and gives the NG bosons~mass.

Similarly, for~the limit of scale and conformal invariance, 
``there are two possibilities: either all particle masses go to
zero, or~there is a massless scalar boson of the NG type that allows
other masses to be non-zero'' \cite{mgm69}. 

The first possibility refers to the WW scaling mode, where 
scale and conformal invariance are manifest. Let
\begin{equation}
D(t) = \int\!d^3x\,{\cal D}_0(t,\bm{x}) \quad\mbox{and}\quad
K_\mu(t) = \int\!d^3x\,{\cal K}_{\mu 0}(t,\bm{x})
\end{equation}
generate scale and special conformal transformations. In~the symmetry
limit, $D$ and $K_\mu$ become time independent, and~their commutators
with the translation and Lorentz generators $P_\mu$ and $M_{\mu\nu}$
simplify, e.g.,
\begin{equation}\label{comm}
\bigl[K_\mu , P_\nu\bigr] = - 2i\bigl(g_{\mu\nu}D + M_{\mu\nu}\bigr) 
 \ , \quad \theta^\lambda_\lambda \to 0 \,.
\end{equation}
Given that $|\text{vac}\rangle$ is the only state annihilated by both
$P_\mu$ and $M_{\mu\nu}$, it follows from (\ref{comm}) that
$K_\mu|\text{vac}\rangle = 0$ implies $D|\text{vac}\rangle = 0$ 
and vice~versa. Conformal invariance of the vacuum state
implies that the theory lies \emph{within} the conformal window:
Green's functions exhibit power-law behavior characteristic of
representations of the conformal group $SO(4,2)$. Dimensional
couplings vanish, e.g.,\ scalar particles decouple from $\theta_{\mu\nu}$:
\begin{equation}\label{nondil}
  \langle\varphi|\theta_{\mu\nu}|\text{vac}\rangle \to 0 \,.
\end{equation}
Particles are massless or do not exist~\cite{georgi}, and~the rest of
the mass spectrum is empty or continuous. Consequently, the~WW-mode
scaling limit is nothing like the real world. Key physical properties
such as a massive spectrum can arise only as \emph{dominant}
contributions from terms in the Hamiltonian which break scale
invariance \emph{explicitly}. 

The other possibility is the NG scaling mode, where there is a
non-compact degeneracy of Poincar\'{e} invariant vacua
\begin{equation}\label{deg}
|\text{vac}\rangle_\rho = \exp\{i\rho D\}|\text{vac}\rangle_{\rho = 0} \,.
\end{equation}
As for the chiral case (\ref{chiral_deg}), these vacua are physically
equivalent; one of them is picked out if a small symmetry breaking
term is added to the Hamiltonian. Equation~(\ref{comm}) remains
valid, so $D|\text{vac}\rangle \not= 0$ implies
$K_\mu|\text{vac}\rangle \not= 0$.  Conformal symmetry is hidden
by the dependence of amplitudes on dimensional
constants such as masses. This is allowed if there is a massless
$0^+$ NG boson $\sigma$ for scale and conformal invariance: the 
\emph{dilaton}%
\footnote{This term was coined in 1969, and~first appeared in print
    in~\cite{isham70}.}.

The key property of a dilaton is that the decay constant $f_\sigma$ in
(\ref{decay}) remains \emph{non-zero} in the scale-symmetric limit:
\begin{equation}\label{dil}
   f_\sigma \not\to 0 \ , \quad \theta^\mu_\mu \to 0 \,.
  \end{equation}
Since $f_\sigma$ has dimensions of mass, amplitudes can depend on
scales in the limit (\ref{dil}). \mbox{Scalar Goldberger–Treiman} relations
of the form (\ref{scalarGT}) become exact, so particles such as
nucleons $N$ can remain massive in a theory with NG-mode scale
invariance. The~pion decay constant $f_\pi$ can also remain non-zero:
the NG mode for conformal invariance is compatible with the NG mode
for chiral invariance~\cite{isham70, ellis70, rjc70, zumino70, pete71, rjc71}.

Evidently, compared with the WW mode, the~NG scaling mode offers the
great advantage that there is a chance that it approximates the real world.
A \emph{small} scale-violating perturbation of the Hamiltonian may be
sufficient to give the dilaton and other NG bosons their observed
small masses and make  small corrections to large masses
in the non-NG particle sector. The~consistency of assuming an NG
mode for scale invariance was confirmed via effective
Lagrangians~\cite{nambu68, mack69, isham70, ellis70, zumino70},
and~by the end of 1970, a~complete understanding had been 
achieved~\cite{zumino70,pete71,coleman}.  
\mbox{However, dilaton phenomenology} at that time did not go far:
the only candidate for $\sigma$ was a vague $0^+$ resonance 
$\epsilon(700)$ which was last listed in the Particle Data Tables 
in 1974. The~main results were for the $\sigma \to \pi\pi$ 
coupling~\cite{ellis70,rjc70}, 
\begin{equation}\label{width}
 f_\sigma g_{\sigma\pi\pi} \simeq - m^2_\sigma \quad
  \mbox{i.e.,\,\ $\sigma$-width $\sim$ $\sigma$-mass,}
\end{equation}
and for the $\sigma \to \gamma\gamma$ coupling due to the
electromagnetic trace anomaly~\cite{rjc72, chan72, chan73}:
\begin{equation}\label{EMtrace}
f_\sigma g_{\sigma\gamma\gamma} \simeq \frac{R e^2}{6\pi^2}
 \ , \quad R = \left.\frac{\sigma(e^+e^- \to \text{hadrons})}{\sigma(e^+e^-
    \to \mu^+\mu^-)}\right|_{\text{high energy}}.
\end{equation}

\section{Gauge~Theories}\label{gauge}

This line of investigation was resumed almost 40 years later, prompted
by a partial-wave analysis~\cite{caprini06} which isolated the broad 
low-mass $0^+$ resonance%
\footnote{A successor $f_0$(400--900) to the dormant $\epsilon(700)$
  resonance was first identified in 1996~\cite{tornq} in the context of
  the linear sigma model. The~key features of the 2006
  analysis~\cite{caprini06}  were its model independence and
  precision, which led to the inclusion of $f_0(500)$ in the 2008
  Particle Data Tables. See~\cite{pelaez16} for an extensive
  review. Our symbol $\sigma$ for the dilaton does \emph{not} mean
  that we rely on the sigma model.} 
$f_0(500)$ at $441 - 272\hsp{0.4}i$ MeV, with~small experimental and
theoretical uncertainties---unlike the $\epsilon(700)$. A~perfect
candidate for the hadronic dilaton of 1968--72 had appeared:
\begin{equation}
\sigma|_{\text{hadronic}} = f_0  \,.
\end{equation}
The mass of $f_0$ is close to $K(495)$ and $\eta(549)$, so 
it makes sense to extend standard chiral $SU(3) \times SU(3)$
perturbation theory $\chi$PT$_3$ to chiral-scale perturbation
theory~\cite{ct1,ct2,ct3} $\chi$PT$_\sigma$, with~NG bosons 
$\pi, K, \bar{K}, \eta, \sigma$ in the combined limit of chiral 
and conformal~symmetry. 

A common question here is: what is so special about the case of 
$N_f = 3$ flavors? 

This has to do with whether good phenomenology results, in~a
first approximation, when a given quark flavor is 
considered to~be
\begin{enumerate}[leftmargin=*,labelsep=4.9mm]
\item heavy enough to be decoupled, or~\item light enough to be part
  of a chiral perturbation theory, or~\item neither. 
\end{enumerate}

The quarks $t$, $b$ and $c$ are far too heavy to belong to
category 2. However, they certainly belong to category 1 if we
restrict ourselves to particle states and operators constructed from
$u,d,s$ quarks and consider amplitudes at energies $\ll m_c$. So, let
us decouple the heavy quarks:
\begin{equation} \label{decouple}
 m_Q \to \infty \ , \quad Q = t,b,c.
\end{equation}

Then, since $u$ and $d$ are much lighter than $s$, it is tempting to
try chiral perturbation theory $\chi$PT$_2$ based on approximate
$SU(2)_L \times SU(2)_R$ symmetry. That gives relatively precise
results (5--10\% accuracy), but~only if the $s$-quark mass $m_s$ is
held fixed. That becomes a problem if we want to make \mbox{a connection}
with scale invariance, because~$\theta^\mu_\mu$ contains a
renormalized version of the mass term $m_s\bar{s}s$. Any attempt to
decouple $s$ by taking the limit $m_s \to \infty$ would be a terrible
approximation, given e.g.,\ the observed $SU(3)_{L+R}$ multiplet
structure of particle~states.

The remaining possibility is that $u$, $d$ and $s$ all belong to
category 2, which corresponds to approximate $SU(3)_L \times SU(3)_R$
symmetry. The~result is $\chi$PT$_3$, a~less precise but adequate
theory $\sim$ 30\% accuracy, apart from difficulties in $0^+$ 
channels due to the low-lying $f_0(500)$ resonance. \mbox{Unitarized chiral}
perturbation theory U$\chi$PT~\cite{truong88, pelaez16} 
is a general dispersive method for dealing with this, \mbox{while 
$\chi$PT$_\sigma$} is a QCD-based effective theory with $f_0$ treated
as a NG boson. These technicalities do not detract from the 
requirement that QCD must make sense in the IR limit that
category-2 masses and all momenta $p_{\text{Operators}/\textsc{ng}}$
for physical operators and NG bosons tend to zero:
\begin{equation} \label{limit}
 p_{\text{Operators}/\textsc{ng}} \to 0 \ , \quad
m_q \to 0 \ , \quad q = u,d,s.
\end{equation}
The key observation is that there is no way of distinguishing this
limit from the infrared limit of the renormalization group (RG) for
gauge theories with massless fermions. We are therefore able to
determine to some extent how the gauge coupling runs in that
limit. That is the benefit of having \emph{no} quarks in category~3.

\begin{figure}[H]
\centering
\includegraphics[width=10 cm]{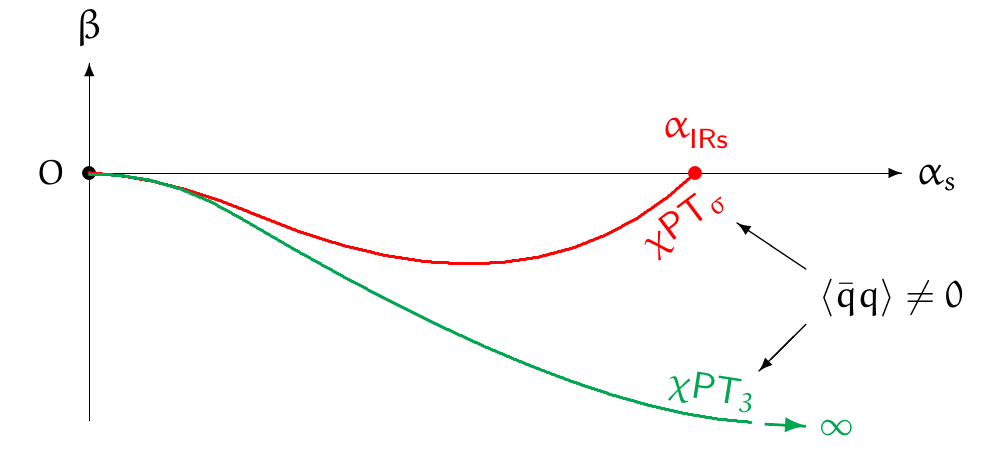}
\caption{Alternatives for the $N_f = 3$ QCD $\beta$-function. (A)
 Conventional $N_f= 3$ soft-meson theory $\chi$PT$_3$ (green curve)
 involves a large breaking of scale invariance at $\alpha_s \sim \infty$
 to ensure that heavy hadrons such as nucleons acquire 
 sufficient mass, but~then that mechanism also generates the mass
 of the $f_0$, which is \emph{not} heavy: $m_{f_0} \sim  m_K \ll
 m_N$. (B) That problem is solved in $\chi$PT$_\sigma$ (red curve): (a)
 the massless dilaton $\sigma$ at ${\alphaIR}_s$ allows nucleons to be
 heavy (Equation (\ref{scalarGT})), and~(b) $\sigma$ becomes the
 pseudodilaton $f_o$ as it acquires a small (mass)$^2$ to first
 order in  $\epsilon_s =  {\alphaIR}_s -  \alpha_s$. Both curves are
 consistent with model-independent U$\chi$PT.
\label{beta}}
\end{figure}

\begin{figure}[H]
\centering
\includegraphics[width=14 cm]{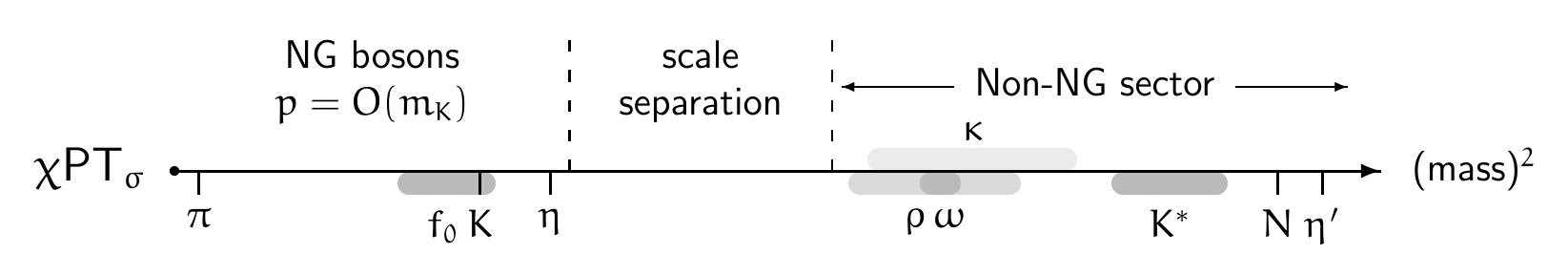}
\caption{The hadronic spectrum below 1 GeV, seen from the point of
  view of $\chi$PT$_\sigma$.  Masses and momenta $p$ of NG bosons,
  including $f_0(500)$, are small relative to scales of the non-NG
  sector. \mbox{Please note} that $\chi$PT$_\sigma$ works only for $N_f = 3$ light
  flavors; there is no analogue of it for $N_f = 2$ because of the
  presence of the $s$ quark: $m_s \gg m_{u,d}$.
\label{spectrum}}
\end{figure}

The result of the decoupling (\ref{decouple}) is QCD for $N_f = 3$
quark flavors $q = u,d,s$.  Since QCD is renormalizable,
there is a trace anomaly~\cite{mink76, adler77, nkn77, collins77}
proportional to the $N_f = 3$ Callan–Symanzik function
$\beta(\alpha_s)$,
\begin{equation} \label{trace}
\left.\theta^\mu_\mu\right|_{\textsc{qcd}}
 =\frac{\beta(\alpha_s)}{4\alpha_{s}} G^{a\mu\nu}G^a_{\mu\nu}
  + \bigl(1 + \gamma_{m}(\alpha_s)\bigr)\!\sum_{q=u,d,s} \!\! m_{q}\bar{q}q
 \,,
\end{equation}
where $\alpha_s$ is the gauge coupling for strong interactions
and $G^a_{\mu\nu}$ is the field-strength tensor. The~theory is
asymptotically free, so $\alpha_s$ increases as low-momentum scales 
are approached. In~the infrared limit~(\ref{limit}), there are two main
possibilities: 
\renewcommand{\labelenumi}{(\Alph{enumi})}  
\begin{enumerate}[leftmargin=*,labelsep=4.9mm]
\item 
  This is the conventional alternative.
  The result is the green curve labelled $\chi$PT$_3$ in
  Figure~\ref{beta}, where $\beta(\alpha_s)$ remains negative and $\alpha_s$
  runs to $+\infty$. In~that limit, the~gluonic part $(\beta/4\alpha_s)
  G^2$ of the trace is still present and breaks conformal invariance
  explicitly. Apart from the massless $0^-$ bosons $\{\pi,K,\eta\}$,
  all hadrons, \emph{including} $f_0(500)$, acquire their mass
  through this mechanism. Please note that the chiral condensate
  $\langle\bar{q}q\rangle_{\text{vac}} \not= 0$ survives in this
  limit.
\item 
  There is an IRFP ${\alphaIR}_s$ at which $\beta$ vanishes and beyond
  which $\alpha_s$ cannot go (the red curve in Figure~\ref{beta}
  labelled $\chi$PT$_\sigma$):
\begin{equation}\label{ir}
 \alpha_s \rightharpoondown {\alphaIR}_s \ , \quad
 \beta({\alphaIR}_s) = 0 \,. 
 \end{equation}
 As a result, both the gluonic and quark-mass terms in
Equation~(\ref{trace}) vanish and the theory becomes conformal invariant:
\begin{equation} \label{QCDconf}
 \left.\theta^\mu_\mu\right|_{\textsc{qcd}} \to\, 0 \,.
 \end{equation}
 Since this is equivalent to the chiral limit (\ref{limit}),  the~
 $0^-$ NG bosons $\pi,K,\eta$ and hence the chiral condensate 
 $\langle\bar{q}q\rangle_{\text{vac}}$ survive, and~so 
 $\langle\bar{q}q\rangle_{\text{vac}}$ acts as a scale condensate.
 That implies the presence of a massless dilaton $\sigma$ at
 the IRFP, which permits all non-NG hadrons to be \emph{massive} 
 in the conformal limit (\ref{QCDconf}). Chiral-scale
 perturbation theory $\chi$PT$_\sigma$ is then a simultaneous 
 expansion about ${\alphaIR}_s$ and in the $u,d,s$ masses. Note 
 the desirable scale separation between the NG-boson sector
$\{\pi,K,\eta,f_0\}$ and heavy hadrons in Figure~\ref{spectrum}. 
\end{enumerate}
\renewcommand{\labelenumi}{\arabic{enumi}.}  
\unskip   

Alternative (A) is possible, but~from the point of view of QCD,
the~small mass of $f_0(500)$ is \mbox{an unexplained} accident. In~a 
nonperturbative setting, the~simplest (and perhaps only) argument
for \mbox{a small} mass is that the theory approximates a symmetry in NG mode.
If $f_0$ is not an NG boson, \mbox{then we} have no symmetry to force its
constituents, $q\bar{q}$ or~\cite{jaffe77, pelaez16}
$q\bar{q}q\bar{q}$, to~be bound together so strongly compared with
other heavy~hadrons.

In alternative (B), the~small $f_0$ mass is due to the approximate
conformal symmetry of $N_f=3$ QCD, together with the small values of
the current-quark masses $m_{u,d,s}$. The~$q\bar{q}$ binding of $f_0$
is similar to that of $\pi,K,\eta$ but in $P$-wave instead of $S$-wave. 

The scale separation shown in Figure~\ref{spectrum} means that an
effective chiral-scale Lagrangian for $\chi$PT$_\sigma$ can be set up
with leading-order (LO) terms given entirely by the tree
approximation. That should be contrasted with conventional
$\chi$PT$_3$, where the tree approximation fails in $0^+$ 
channels and the LO must be patched up with unitarized $\pi, K,
\eta$ loops.  In~$\chi$PT$_\sigma$, the~role of unitarization is to
patch up \emph{next-to-LO} $\pi,K,\eta,\sigma$ loop diagrams; so far,
little has been done in that~regard. 

Concerns about this scheme typically run along the following lines:
\renewcommand{\labelenumi}{(\alph{enumi})} 
\begin{enumerate}[leftmargin=*,labelsep=4.9mm]  
\item 
\emph{IRFP's outside the conformal window are not taken seriously
 in the literature: they do not exist either in principle or on the lattice.} 
 Questions of principle and evidence from the lattice are analyzed in
 Sections~\ref{irfp} and \ref{renorm} respectively. 
\item 
\emph{There are no light dilatons in gauge
 theories~\cite{hold87,hold88}.} These claims are made for a type-III 
 definition of ``dilaton'', to~be discussed further in
 Section~\ref{competing} below. They do not affect the identification
 above of the $f_0(500)$ as a genuine dilaton $\sigma$ for QCD.
\item
\emph{There is no light dilaton in hadronic physics because there is
 no scalar particle nearly degenerate with pions.} This overlooks the
 role of $m_s \gg m_{u,d}$, and~comes from TC literature, where all
 chiral NG bosons are called ''technipions'' and none ``technikaons''
 or ``technietas''.  
\item  
\emph{A light dilaton is not seen for $N_f=4$.} This refers to the
lattice study~\cite{appel19} closest to the relevant case $N_f = 3$. See
Section~\ref{renorm}.  
\item
\emph{There may be an IRFP for \mbox{$N_f=2$} which would, in~analogy with
 the case \mbox{$N_f=3$}, produce a spin-$0^+$ particle with mass $O(m_\pi)$,
 contrary to experiment.} An IRFP at \mbox{$N_f=2$} is not excluded
 but, as~noted above, a~connection with scale invariance can be
 obtained only by decoupling the $s$ quark, and~$m_s \sim \infty$ is a
 very bad approximation. The~argument works only for \mbox{$N_f=3$}.
\end{enumerate}
\renewcommand{\labelenumi}{\arabic{enumi}.}  

Continuing with the case \mbox{$N_f = 3$}, let us consider the LO
approximation for
\begin{equation}
2m_\sigma^2 
= \langle\sigma\bigl|\theta^\mu_\mu\bigr|\sigma\rangle
 \simeq \frac{{\alphaIR}_s - \alpha_s}{4{\alphaIR}_s}
  \beta'_{\textsc{qcd}}\langle\sigma|G^2|\sigma\rangle 
     + \bigl(1+\gamma_m({\alphaIR}_s)\bigr)\!
\sum_{q=u,d,s}\! m_q  \langle\sigma|\bar{q}q|\sigma\rangle \,.
\end{equation}
Here $\beta'_{\textsc{qcd}} \geqslant 0$ is the slope at the IRFP of the
red curve in Figure~\ref{beta}. The~optimal space-like scale $-m^2$ at
which $\alpha_s = \alpha_s(-m^2)$ should be evaluated is determined
by how close to the limit (\ref{limit}) it is possible to go in the
real world. Soft-meson theorems for approximate $SU(3) \times SU(3)$ 
involve $O(m_K)$ extrapolations in NG-boson masses and momenta%
\footnote{Including pion momenta, as~in $\eta \to 3\pi$
  decay. Distinguish $\pi\pi  \to \pi\pi$ for $O(m_K)$ momenta from   
  the same process for $O(m_\pi)$ momenta, where $\chi$PT$_2$, 
  a~different theory, is applicable. See Footnote 7 and Figure~4 
  in~\cite{ct2}.},
so we take $m \sim m_K$. Therefore, relative to a QCD large scale
such as $M_N$, effects due to
\begin{equation} \label{near}
\epsilon_s 
= {\alphaIR}_s - \alpha_s 
 \ \simeq \alpha_s(0) - \alpha_s\bigl(-m_K^2\bigr)
= O\bigl(m_K^2/m_N^2\bigr)
\end{equation} 
are similar in magnitude to those of $m_s$. That is why $f_0(500)$ is
as \emph{light} as $K$ and $\eta$ (Figure \ref{spectrum}). 

The physical consequences of having an IRFP at ${\alphaIR}_s$ can be
seen in low-energy mesonic processes. The~results are encoded in a
chiral-scale Lagrangian $\left.{\cal L}_{\text{eff}}\right|_{\textsc{qcd}}$ for
$\chi$PT$_\sigma$~\cite{ct1,ct2}.  
The~formalism is entirely standard, having been invented in 
1969--1970~\cite{salam69,isham70,ellis70,zumino70}, 
so will not be repeated in full here. Under~global conformal
transformations $x \to x'$, the~Goldstone field $\sigma$ is translated
by a constant: 
\begin{equation}
\sigma\ \to\ \sigma -  (f_\sigma /4)\ln|\det(\del x'/\del x)| \,.
\end{equation}
Then $\exp(\sigma d/f_\sigma)$ transforms covariantly with dimension
$d$ and can be used to adjust the dimensions of terms in an effective
Lagrangian. For~example, the~mass term $M_N\Nbar N$ for
a nucleon can be converted into a scale-invariant potential $V_{\text{inv}}$,
\begin{equation}
M_N \Nbar N\ \longrightarrow\ 
V_{\text{inv}} =  M_N e^{\hsp{0.1}\sigma/f_\sigma}\Nbar N 
 = M_N \Nbar N \bigl\{1 + \sigma/f_\sigma + \ldots\bigr\} \,,
\end{equation}
from which the scalar Goldberger–Treiman relation (\ref{scalarGT}) may
be deduced. So, any effective Lagrangian can be made conformal
invariant by introducing a suitable dependence on $\sigma$. The~key
point is that this produces amplitudes which depend on scales such as
$M_N$, $f_\pi$ and $\Lambda_{\textsc{qcd}}$ in the limit of
\emph{exact} conformal invariance, i.e.,\ \emph{at the IRFP
$\alphaIR$.} 

The same technique is applied to a chiral Lagrangian by noting
that the unitary matrix field $U$ for chiral NG bosons has dimension
0.  For~example, the~Lagrangian
\begin{equation} \label{model}
{\cal L}_{\text{inv}}
 = \frac{1}{2} \del^\mu\sigma \del_\mu\sigma e^{2\sigma/f_\sigma} 
   + \frac{1}{4} f^2_\pi \mbox{Tr}\, \del^\mu U \del_\mu U^\dagger
        e^{2\sigma/f_\sigma}   
\end{equation}
has dimension-4: both chiral and scale invariance are preserved.
The term $|\del\vec{\pi}|^2\sigma/f_\sigma$ obtained from
Equation~(\ref{model}) corresponds to the result (\ref{width}) for
the dilaton~width. 

The effective Lagrangian for $\chi$PT$_\sigma$ generalizes
(\ref{model}) to include all possible LO terms consistent with the
conformal and chiral $SU(3)_L \times SU(3)_R$ properties of 
QCD for small values of $m_{u,d,s}$ and $\epsilon_s$. The~most
important result is that the $\Delta I = 1/2$ rule for nonleptonic
kaon decays is a consequence of broken scale and chiral invariance. 
Equation~(\ref{width}) remains valid~\cite{ct3}, while $R$ in
Equation~(\ref{EMtrace}) is replaced by the high-energy ratio
$R_{\textsc{ir}}$ for the scale-invariant theory at $\alphaIR$.

Crawling TC~\cite{crawl} is the most recent application of the idea of
an NG mode at the IRFP of a gauge theory. It adopts the standard TC
viewpoint~\cite{wein76,wein79,suss79} that the Higgs mechanism is the
dynamical effect of a gauge theory which resembles QCD, with~a TC
coupling $\alpha$ which is nonperturbative at scales of \mbox{a few} TeV.
Where it differs from other TC theories is that, in~analogy with (\ref{ir}),
the TC gauge coupling runs to an infrared fixed point $\alphaIR$ with
conformal invariance in NG mode. So, at~$\alphaIR$, there is a massless
dilaton---a feature unique to crawling TC. At~energies much less than
a TeV, $\alpha$ sits just below $\alphaIR$ and the dilaton acquires a
mass $\ll$ TeV. It makes sense to identify this massive $\sigma$
particle with the mass \mbox{125 GeV} Higgs~boson.

The characteristic feature of crawling TC is its dependence on the 
slope $\beta'$ of the TC $\beta$ function at the fixed point:
\begin{equation}
\beta' = \left.\frac{d\beta}{d\alpha}\right|_{\alpha = \alphaIR} 
           > 0 \,.
\end{equation}
In particular, the~Higgs potential in leading order is a
nonpolynomial function
\begin{equation}\label{pot}
V(h) = \frac{M_{\sigma}^2F_{\sigma}^2}{\beta'}
\left[-\frac{1}{4}
\left(1+\frac{h}{F_\sigma}\right)^4\! + \frac{1}{4+\beta'}
\left(1+\frac{h}{F_\sigma}\right)^{4+\beta'} \!+ \frac{\beta'}{4(4+\beta')}\right] \,,
\end{equation}
where $h = h(x)$ is a fluctuating Higgs field, $F_\sigma$ is the 
TC analogue of $f_\sigma$, and~$M_\sigma$ is identified as the Higgs
boson mass $m_h$. 

\section{Comparison of Crawling and Walking~TC}\label{competing}

While writing~\cite{crawl}, we became aware that the 1968--70 concepts 
of ``dilaton'' and ``spontaneous breaking of conformal invariance'',
on~which our work relies, have lost their original meaning
(Section~\ref{intro}). 
Most of the thousands of papers on the subject written since 1972 do
not recognize the type-I concept that \emph{scale-dependent}
amplitudes can occur in the limit of conformal invariance. For~most
authors, the~label ``dilaton''  is just a fancy name for a scalar
field appearing in a conformal theory. \mbox{That has} led to a lack of
clarity between competing~concepts.

Most definitions of ``dilaton'' on the Internet are of type-II: they
refer to a scalar component of the gravitational field.  There is now
a vast literature on this. The~term was first used in that context in 
1971~\cite{fujii71}. At~the time, it drew the remark~\cite{mgm71} (quoted
in~\cite{rjc14}) that ``Brans–Dickeon'' would be \mbox{a better~name}. 

There is a third meaning for ``dilaton'' (type-III), also with an
extensive literature, \mbox{which unfortunately} contradicts the 1968--70
definition reviewed above. This re-working of the subject started in
1976: 
\begin{enumerate}[leftmargin=*,labelsep=4.9mm]
\item Fubini~\cite{fubini76} noted problems with the conformal 
 NG mode for $\lambda \phi^4$ theory which subsequent authors
 incorrectly interpreted as an inconsistency of type-I theories in
 general; see Section~\ref{scales} below.
\item Gildener and Weinberg (GW) \cite{gild76} introduced the concept
 of a spin-$0^+$ ``scalon''  associated with \mbox{a flat} direction of
 the potential of a massless gauge theory in the tree approximation.
 \mbox{Scale invariance} is broken \emph{explicitly} by one-loop corrections of
 the Coleman–Weinberg (CW) \cite{cw73} type. The~analysis is entirely 
 consistent, except~for a remark that the result is an example of a
``spontaneous breaking'' of scale invariance%
\footnote{Coleman and E.~Weinberg~\cite{cw73} stick to the textbook
  definition of the term ``spontaneous'', i.e.,\ for breaking which is
  \emph{not} explicit, and~apply it \emph{only} to the breaking of
  chiral invariance. In~footnote 8, they note that scale invariance
  is broken explicitly by the one-loop trace anomaly. \label{CW}}.
That is not so: the tree approximation is scale-free by construction,
so the invariance is realized in the WW mode. In~that limit,
the~``scalon'' is massless but is not a genuine dilaton because it
lacks a decay constant connecting $\theta_{\mu\nu}$ to the vacuum. All
breaking of conformal invariance is \emph{explicit}: the one-loop
corrections violate scale invariance of the Hamiltonian.
\end{enumerate}

As reported in Section~\ref{intro}, the~GW scalon became the type-III
``dilaton'' of walking 
TC~\cite{bardeen86,yam86,hold87,hold88,appel10,yam11,yam14,golt16,golt18}
and conformal potentials deformed by the CW
mechanism~\cite{meiss07,chang07,foot07,gold08,vecc10,bell13,bell14,cora13}.  
Exact conformal invariance is clearly in the WW mode, e.g.,~within the
conformal window for walking TC, yet  the 
``breaking'' of the symmetry is said to be both ``explicit'' and
``spontaneous''. Other versions of this contradiction are that
``approximate conformal invariance is spontaneously broken'', or~that
the breaking ``triggers''  scale generation ``spontaneously''. 
 
A general definition for the type-III dilaton $\varphi$ for walking TC
and CW-deformed potentials is as follows. It is a $0^+$ particle in a
theory which approximates a system with \emph{exact} conformal
invariance in the WW mode,
\begin{equation}\label{ww}
D|\text{vac}\rangle = 0 = K_\mu |\text{vac}\rangle 
 \ , \quad \theta^\mu_\mu \to 0 \,,
\end{equation}  
and obeys Equation~(\ref{nondil}). All scales, large and small, are
``triggered''  when the Hamiltonian%
\footnote{In walking TC, the~decomposition ${\cal H} = {\cal H}_0 +
  \delta{\cal H}$ is often not considered explicitly. Such theories involve
  extrapolations in $N_f$, with~an understanding that the extra
  flavor fields are almost decoupled.}
is perturbed by a term which
breaks conformal invariance \emph{explicitly}. These large scales
include a fermion condensate 
$\langle\overline{\psi}\psi\rangle_{\text{vac}}$ and hence chiral NG
bosons. 

Walking TC assumes that all infrared fixed points lie within the 
conformal window, where deep infrared dynamics is scale-free and 
Green's functions exhibit the power-law scaling expected for the WW
mode. The~gauge coupling $\alpha$ for a theory just outside the
conformal window is supposed to walk slowly when it passes the IRFP 
$\alphaWW$ of a theory just inside the window. The~result is then \mbox{a
small} $\beta$-function which (it is hoped) can be held responsible for
small-scale effects such as the mass of the Higgs boson. Physics
outside the conformal window is vastly different from physics inside,
so there must be a discontinuity or phase transition in $N_f$
at a sill~\cite{appel96,appel97,LSD10} produced by a term $\delta{\cal
  H}$ in $\theta_{00}$ which breaks conformal symmetry
explicitly. Despite being proportional to $\beta$, $\delta{\cal H}$
must produce \mbox{effects $\sim$ several TeV}, such as
$\langle\overline{\psi}\psi\rangle_\text{vac}$. 
 
Therefore, even though $\theta^\mu_\mu$ is 
formally small, its effects are $\sim$ a few TeV, as~foreshadowed in
Section~\ref{intro} and in general remarks below Equation~(\ref{nondil}). 
So it is hard to argue that the sill produces a small-mass
Higgs boson. That can be done only if an explicit model for the sill
can be formulated with unusual properties. The~model would have to
specify a large-scale mechanism for the gauge theory to~produce the
chiral condensate $\langle\overline{\psi}\psi\rangle_{\text{vac}}$
\emph{without} affecting the mass of the $0^+$ boson. Early attempts
in~that direction~\cite{bardeen86, hold87,hold88} came to the conclusion
that this is not possible: type-III dilatons are heavy%
\footnote{The analysis does not appear to depend on their having an
  ultraviolet (UV) fixed point instead of an IRFP.}.
Of course, the~self-consistency of these gauge-theory
models for fermion condensation is far from obvious, but~that does
not mean that the difficulty can be circumvented by assuming 
(as in~\cite{gold08}) that a type-I dilaton Lagrangian from 
1970~\cite{isham70,ellis70,zumino70} may be valid below the sill but not
above it. For~that to be convincing, the~self-consistent model for the sill
would have to produce \mbox{a dilaton-like~Lagrangian}. 

\begin{figure}[H]
\centering
\includegraphics[width=15 cm]{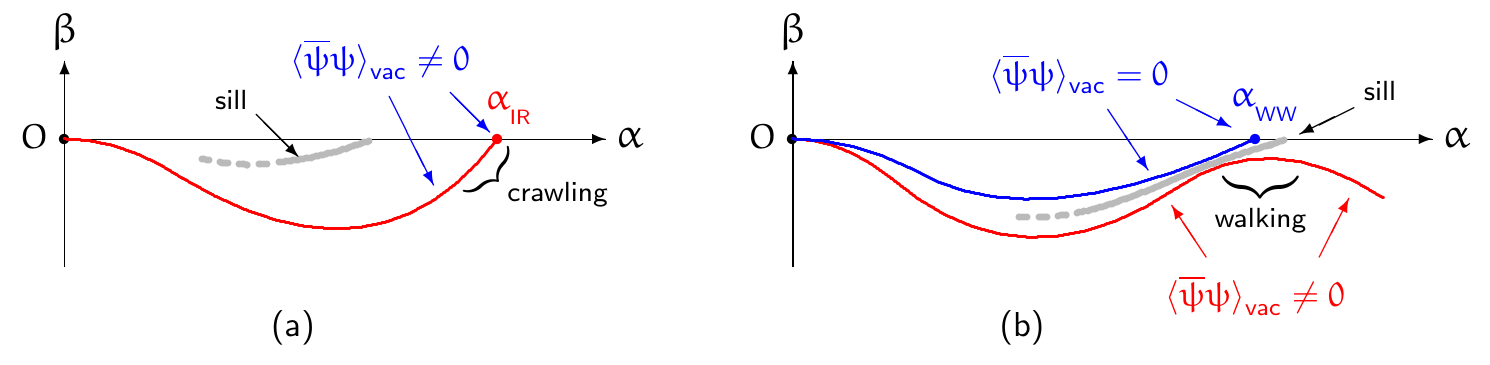}
\caption{Graphs of the TC $\beta$ function in $SU(3)$ gauge theories  
 with $N_f$ Dirac flavors for (\textbf{a}) crawling TC and  (\textbf{b})
 walking TC. On~each graph, the~sill of the conformal window at 
 $N_f^\mathrm{c}$ flavors is shown as a thick gray line.  
 In~(\textbf{a}), there is a genuine massless dilaton at $\alphaIR$  
 outside the conformal window ($N_f \leqslant N_f^\mathrm{c}$), so
 large scales are  possible at that point. Small-scale corrections
 such as $m_h = 125$ GeV for the Higgs boson mass occur for $\alpha
 \lesssim \alphaIR$.  
 In (\textbf{b}), the~physical theory below the sill (red  line, $N_f
 \leqslant N_f^\mathrm{c}$) is scale-dependent and lacks an IRFP. 
 The theory above the sill and hence inside the conformal
 window (blue line, $N_f^\mathrm{c} <  N_f \leqslant 16$) is
 scale-free and has an IRFP  $\alpha_{{}_{\text{WW}}}$ with conformal
 invariance in WW  mode. The~sill generates all scales in the physical
 theory, both large and small,  no matter how closely the red line in
 the walking region approaches $\alpha_{{}_{WW}}$, with~the walking
 $\alpha$ evaluated at a space-like scale $\sim
 -\Lambda^2_{\textsc{tc}}$. Therefore, a~type-III ``dilaton'' is 
 unlikely to be light. 
\label{tc}}
\end{figure}

This should be contrasted with crawling TC, which relies on a single
assumption that there is \mbox{an IRFP} $\alphaIR$ in the NG mode for
conformal invariance. Support for this comes from evidence noted in
Section~\ref{gauge} for the analogue theory~\cite{ct1,ct2,ct3} for QCD.
Assumptions about the detailed dynamics of the sill are not needed;
indeed, the~sill plays no role in the extrapolation from $\alphaIR$ to 
$\alpha$. That extrapolation accounts for small-scale corrections to
the scale set at $\alphaIR$, including the mass of the Higgs boson
\begin{equation}
m_h = O(\epsilon) \ , \quad
\epsilon = \alphaIR - \alpha \gtrsim 0 \,.
\end{equation}
See~\cite{crawl} for an  explicit $O(\epsilon)$ formula for $m_h$ in
terms of the gluon condensate at $\alphaIR$. A~diagrammatic comparison
of the two theories is shown in Figure~\ref{tc}.

At what scale should $\alpha$ be evaluated? Unlike QCD, where the 
relevant scale in (\ref{near}) was found to be set by the heaviest
light quark $s$, TC theory is a gauge theory with massless fermions
with no obvious analogue of \emph{approximate} chiral $SU(3)_L \times
SU(3)_R$ symmetry. However, TC is in some way perturbed by the 
electroweak theory responsible for the $W^\pm$ and $Z^0$ bosons
and lighter non-TC particles. So, it is tempting to suppose that the
optimal scale for $\alpha$ is set by the largest small (non-TC) scale
available, i.e.,\ $M_{W,Z}$:
\begin{equation}
\epsilon \simeq \alpha(0) - \alpha(-M^2_{W,Z}) 
  = O\bigl(M^2_{W,Z} \bigl/\Lambda^2_{\textsc{tc}}\bigr) \,.
\end{equation}
Here $\Lambda_{\textsc{tc}}$ is a typical nonperturbative TC scale 
$\simeq$ a few TeV. This would explain why the Higgs boson is almost
as \emph{light} as $W^\pm$ and $Z$.

Theoretical support for this outcome requires the construction of a
fully unified gauge theory which combines the Standard Model with 
TC. That deserves further~investigation.
 

\section{Scale Dependence in the Conformal~Limit}
\label{scales}

Evidently the proposition that amplitudes at $\alphaIR$ can be 
scale-dependent requires further explanation. In~the limit of exact
conformal invariance, (a) is scale dependence of the ground state
generally possible, and~(b) can it occur at an IRFP of a massless 
gauge theory  (Section \ref{irfp} below)?

The argument against (a) typically refers to Fubini~\cite{fubini76} and
runs as follows~\cite{bell14}: 
``if a theory is exactly conformal,  it either does not break scale
invariance, or~the breaking scale is arbitrary (a flat direction).''
In effect, it is being argued that the NG mode for \emph{exact}
conformal invariance with a type-I dilaton is absolutely impossible.
That cannot be~so:
\begin{enumerate}[leftmargin=*,labelsep=4.9mm]
\item Fubini's analysis is restricted to $\lambda\phi^4$ theories
 and therefore does not constitute a general proof that strict conformal
 invariance must be manifest, i.e.,~in WW mode. To~obtain the NG mode
 for conformal invariance, simply omit the $\phi^4$ term and add other
 invariants to $\frac{1}{2}(\del\phi)^2$ such as couplings to chiral NG
 bosons or (say) the 4-point self-interaction
\begin{equation}
{\cal L}_{\text{4-pt.}} = \kappa\bigl(\del\phi\bigl/\phi\bigr)^4 \ ,
  \quad  \kappa = \mbox{const.},
\end{equation}
and~\cite{crawl} \emph{constrain} $\phi$, e.g.,\ to a half line
\begin{equation}\label{constraint}
\phi > \mbox{ const.} 
\end{equation}
\item As noted in~\cite{crawl}, Fubini's conclusion was anticipated
 in 1970 by Zumino (page 472 of~\cite{zumino70}), \mbox{who observed} that a
 dilaton Lagrangian is consistent only if the quartic term
 \emph{vanishes} in the \mbox{conformal  limit}:
\begin{equation} \label{bruno}
\lambda = O(\epsilon) \ , \quad \epsilon \to 0  \,.
\end{equation}
 Here $\epsilon$ is a measure of the explicit breaking of conformal
 symmetry. Equation~(\ref{bruno}) reflects the fact that, like other
 genuine NG bosons, type-I dilatons for $\epsilon \to 0$ are massless
 and cannot self-interact at zero momentum: they correspond to a flat
 direction of the dilaton potential. 
\item All dilaton Lagrangians from 1968--1670 which obey Zumino's
 rule (\ref{bruno}) are counterexamples~\mbox{\cite{nambu68,mack68,mack69,isham70,ellis70,zumino70}}:
 they exist in the limit  of exact conformal symmetry and produce
 amplitudes which depend on a \emph{non}-arbitrary scale,  the~dilaton
 decay constant $f_\sigma$ of Equations~(\ref{decay}) and  (\ref{dil}).
 All except~\cite{nambu68} allow chiral condensates to exist in
 the conformal limit $\epsilon \to 0$.
\item The ``flat direction'' is \emph{not} associated with a continuum 
 of scales. Instead, it corresponds to the continuum of degenerate
 vacuum states (\ref{deg}).
\end{enumerate}
The quote continues: ``Thus an explicit breaking must be present to
trigger and stabilize the spontaneous breaking of scale~invariance.''
\begin{enumerate}[leftmargin=*,labelsep=4.9mm]\setcounter{enumi}{4}
\item Again, the~effective Lagrangians above are counterexamples.
 A  tiny $0(\epsilon)$ scale-violating perturbation 
 $\delta{\cal H}_\text{tiny}$ can pick out one of the degenerate vacua
 (stabilization) and produce \emph{tiny} corrections to the
 scale-dependent amplitudes and masses  of the type-I theory at
 $\epsilon = 0$. 
\item Implicit in this quote is the type-III assumption that there are
 no scales in the $\epsilon = 0$ theory, so it is necessary to
 have a large discontinuity appear ``spontaneously'' at a small
 or infinitesimal value of $\epsilon \not= 0$ to produce large
 scales. If~the $\epsilon =  0$ theory is in the WW scaling mode,
 it does not have scale-degenerate vacua, so there is nothing to stabilize.
\item The large discontinuity is a problem for type-III phenomenology,
 because $\theta^\mu_\mu \sim 0$ is such a bad approximation. 
\end{enumerate}

A formal argument that ``the breaking scale is arbitrary'' in a
conformal invariant theory was first given by Wess~\cite{wess60}.
It is most simply derived from the identity~\cite{mack69}
\begin{equation}
e^{iD\rho} P^2 e^{-iD\rho} = e^{2\rho} P^2 \ , \quad 
   \theta^\mu_\mu \to 0 \,,        
\end{equation}
which implies that mass-$\cal M$ eigenstates $|{\cal M}\rangle$ obey the
relation
\begin{equation}
|\hsp{0.3}e^{\hsp{0.1}\rho}\!{\cal M}\rangle = e^{iD\rho}|{\cal M}\rangle  \,.
\end{equation} 
That implies a spectrum of zero-mass particles, or~a continuum 
$0 \leqslant {\cal M} < \infty$, or~both---provided that the ground
state is unique (WW mode of conformal invariance).

However, for~vacua (\ref{deg}) degenerate under scale transformations, 
this conclusion is not valid because states related by $e^{iD\rho}$
belong to different worlds, $W$ and $W'$. A~discrete scale $\cal M$
can exist in $W$ and correspond to a discrete scale ${\cal M}'$ in
$W'$. Since dimensional units are also scaled up or down in the same
way, e.g.,
\begin{equation}
\mbox{GeV} \to \mbox{GeV}' = e^{\hsp{0.1}\rho}\,\mbox{GeV} \,,
\end{equation}
experimental data in $W$ and $W'$ are identical. Therefore these
worlds are physically equivalent, as~for any other symmetry in the
NG mode. See Appendix D of~\cite{crawl} for~details.

Sometimes type-I dilaton Lagrangians are written in a form such that
scales do not appear explicitly in the conformal limit. That happens
when all fields are chosen to transform homogeneously under scale
transformations. The~result is a polynomial Lagrangian with
dimensionless coupling constants which is easily confused
with the conformal WW mode considered by Gildener and
Weinberg~\cite{gild76}. The~difference for the NG mode is that the
scale may be hidden in a constraint like (\ref{constraint}) which must be
implemented nonlinearly. The~simplest example is the constraint $\phi
> 0$ which is not changed by scale transformations and seems to have
no scale dependence. However, to~implement it, a~scale must be
introduced, as~is evident from the mapping~\cite{salam69}
\begin{equation}\label{a}
\phi = f_\sigma \exp\bigl(\sigma\bigl/f_\sigma\bigr) > 0
\end{equation}
from the unconstrained Goldstone field $\sigma$. 

Flat directions for conformal invariant Lagrangians are also possible for
type-III theories, as~noted by Gildener and Weinberg~\cite{gild76},
but~they do \emph{not} correspond to the vacuum degeneracy (\ref{deg}) 
because \mbox{a type-III} vacuum state is conformal
invariant. Instead, the~flat direction corresponds to
\mbox{field-translation invariance} 
\begin{equation}
\phi(x) \to \phi(x) + c \quad\mbox{for}\quad -\infty < c < \infty 
\,, \ \ c =\ \text{const.},
\end{equation}
which forbids definitions like (\ref{a}) that introduce a
scale. In~particular, Equation~(\ref{a}) cannot be used 
above the sill of the conformal~window.

\section{Scale Dependence at an~IRFP}
\label{irfp}

There is an extensive literature on IRFP's, but~in almost all of it,
``conformality'' (a lack of scale dependence at IRFP's) is accepted
without question. This may be~because:
\begin{enumerate}[leftmargin=*,labelsep=4.9mm]
\item The initial work~\cite{caswell74} was perturbative with a
 scale-free IRFP. That implied manifest chiral symmetry, so an
 IRFP of that type would presumably be close to a discontinuous
 transition to a phase where fermions can condense~\cite{banks82}. 
 That became the model for walking TC.
\item It is relatively easy to find scale-free IRFPs on the
 lattice: Green's functions exhibit power-law behavior in the 
 conformal window. That does \emph{not} test the possibility of
 IRFP's outside the conformal window (Section \ref{renorm}).
 \item There is a belief that dimensional transmutation, which 
 produces nonperturbative scales like  $\Lambda_\textsc{qcd}$ or
 $\Lambda_\textsc{tc}$, implies $\theta^\mu_\mu \not= 0$. If~true,
 that would exclude scale dependence at IRFPs. 
\end{enumerate}

When using the term ``dimensional transmutation'',
care must be exercised not to conflate two distinct concepts:
\renewcommand{\labelenumi}{(\alph{enumi})}  
\begin{enumerate}[leftmargin=*,labelsep=4.9mm] 
\item RG-invariant scales $\cal M$ induced by the renormalization scale
 $\mu$ of $\alpha$,
\begin{equation}
{\cal M} = \mu \exp \bigg\{-\int^\alpha_{\kappa_{\cal M}} 
                  dx\bigl/\beta(x)\bigg\} \,,
\quad 0 < \kappa_{\cal M} < \alphaIR \,,
\label{M_inv}
\end{equation}
where $\kappa_{\cal M}$ is a dimensionless constant that depends on
$\cal M$ but not on $\alpha$ or $\mu$. Examples of $\cal M$ for
massless $N_f=3$ QCD are non-NG hadron masses such as $M_N$ and
dimensional constants like $f_\sigma , f_\pi$ and $\Lambda_\textsc{qcd}$,
and for TC, their counterparts such as $F_\sigma , F_\pi$ and
$\Lambda_\textsc{tc}$. 
\item The trace anomaly which, if~present, is also induced by $\mu$.
\end{enumerate}
\renewcommand{\labelenumi}{\arabic{enumi}.}  
If (a) and (b) are conflated, the~idea that dimensional transmutation
and hence scale dependence may occur at an IRFP looks like an absolute
contradiction. This confusion in terminology has arisen because the
original CW analysis was performed at \emph{one-loop}
order\footref{CW}. In~that order, an~IRFP cannot occur, so there was
no need to distinguish (a) and (b).

There is no proof%
\footnote{See Section~2 of~\cite{crawl}, especially the text below
  Equation~(27).}
that ${\cal M} \not= 0$ implies $\beta \not= 0$. 
As $\alpha$ moves from the perturbative region into the hadronization
region and then beyond into the infrared region, it is hard to argue
that all RG-invariants $\cal M$ suddenly turn themselves off when a
nonperturbative IRFP is encountered.   So there is a theoretical
possibility that dimensional transmutation \emph{in the sense of (a)}
may occur at an IRFP $\alphaIR$. It should be~investigated.

\begin{figure}[H]  
\centering
\includegraphics[width=10 cm]{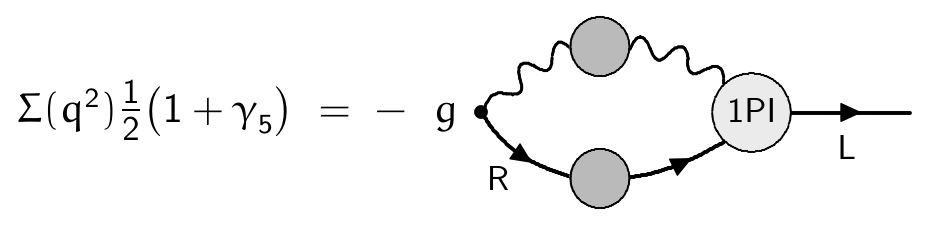}
\caption{ \label{fermion}
Self-consistent condition for the fermion self-energy $\Sigma$,
with $\alpha = g^2/(4\pi)$. The gauge-boson and fermion propagators
are fully dressed, while $\Sigma$ and the vertex labelled 1PI are
one-particle-irreducible.} 
\end{figure}

Fermion condensation is a special case of this. Figure~\ref{fermion}
shows the standard condition for the self-energy $\Sigma$ implied by
the Schwinger-Dyson equation for the fermion propagator. If~a non-zero
solution for $\Sigma$ and hence $\langle\bar{\psi}\psi\rangle_{\text{vac}}$  
exists for finite values of the fermion-gluon coupling constant
$g$, \mbox{why should} this result not be valid at the value $g_{\textsc{ir}}$
corresponding to the IRFP $\alphaIR = g^2_{\textsc{ir}}/(4\pi)$?  

To my knowledge, the~only argument against this conclusion is a claim 
that the dynamical mass acquired by fermions due to the condensate
would cause them to become relatively heavy at low energies and so
decouple in the infrared limit. The~trouble with that is evident from
Figure~\ref{beta} for QCD with $N_f=3$ flavors. Decoupling of
$u,d,s$ would imply that the quark condensate $\langle\bar{q}q\rangle_{\text{vac}}$
and hence $\pi,K,\eta$ decouple in the infrared limit (\ref{limit}). That
would destroy chiral $SU(3)_L \times SU(3)_R$ perturbation theory and
spell the end of QCD, irrespective of whether an IRFP exists or~not.

The hole in this decoupling argument was examined at length in
Appendix A of~\cite{crawl}. It has to do with the distinction between
current and constituent quarks. Current quarks refer to the $u,d,s$
fields in the QCD Lagrangian, with~small ``current-quark'' masses
$m_{u,d,s}$ which govern the masses of $\pi,K,\eta$. Constituent or
``dressed'' quarks have large masses  $M_{u,d} \sim$ 300 MeV and 
\mbox{$M_s \sim$ 450 MeV} in \mbox{a quantum-mechanical} Hamiltonian
$H$ which reproduces the spectrum of non-NG hadrons, e.g.,~$M_N = 2M_u
+ M_d$. The~constituent masses are  ``dynamical'' because $H$ is (presumably)
the result of integrating out the NG-boson sector and so has its scale
set by $\langle\bar{q}q\rangle_{\text{vac}}$. 

Obviously, the~constituent masses cannot be regarded as masses in the 
QCD Lagrangian because that would prevent the chiral limit being
taken, and~pions would have mass $\sim 2M_{u,d}$. So, one would expect
the Appelquist–Carazzone theorem~\cite{AC} to apply only to heavy
current-quark masses such as $m_{t,b,c}$, and~not to $M_{u,d,s}$. That
is the result found in~\cite{crawl}.

Evidently items 1--3 are assumptions characteristic of type-III
theories. For~the type-I theories $\chi$PT$_\sigma$ and crawling TC,
the problem is to find a satisfactory replacement for item 2.
If~scales are present at $\alphaIR$, Green's functions do \emph{not}
exhibit power-law behavior; rather, they behave much like
amplitudes observed in the real~world.

\section{Nonperturbative Tests of Type-I~Theories}
\label{renorm}

The obvious tactic is to define $\alpha$ non-perturbatively outside
the conformal window and see if it stops increasing as the infrared
limit is approached. There are two~difficulties:
\begin{enumerate}[leftmargin=*,labelsep=4.9mm]
\item A true analogue of the Gell-Mann--Low function $\psi(x)$
  for quantum electrodynamics~\cite{GML} is assumed to exist for
  non-Abelian gauge theories but is yet to be identified.
  Prescriptions for the running coupling exist beyond perturbation
  theory~\cite{brodsky}, but~there is a danger that their properties
  are artefacts of their definition. We have no analytic proof that
  any of them runs monotonically and provides an unbiased test of
  whether the dynamics chooses to have an IRFP or not. The~method of
  effective charges~\cite{grun84,lusch10,kuti12} is nonperturbative,
  but there are as many definitions as there are physical processes,
  and it is not obvious which of them has the desired properties all
  the way to the far infrared.
\item Lattice studies~\cite{FLAG,Brida} feature precise measurements of 
 $\Lambda_{\textsc{qcd}}$ in UV logarithms ($\beta \sim$ perturbative)  
 and clear evidence for hadronization and quark condensation at
 intermediate energies.  \mbox{However, for~small} $N_f$ values such as $N_f=3$,
 it is hard to reach the infrared region far below the non-NG hadronic
 spectrum. 
\end{enumerate}

A less ambitious procedure is to look for a light scalar particle in
the particle spectrum. The~problem then is to decide whether this is
evidence for a type-III or a type-I~theory.

In the context of walking TC, the~most interesting cases are those
just under the sill of the conformal window. Evidence for a light
scalar particle almost degenerate with technipions has been found in
lattice data for $SU(3)$ with $N_f = 8$ Dirac fermions in the
fundamental representation~\cite{aoki14,aoki17,appel16,appel19}  and
two Dirac fermions in the sextet
representation~\cite{fodor14}. In~each case, the~particle is
identified as a~``dilaton''. In~this 
type-III interpretation, the~small mass is considered to be due to
$\alpha$ being close to \mbox{a WW-mode} fixed point $\alphaWW$  
just inside the conformal~window. 

However, as~explained above, that involves unlikely assumptions about  
the dual character of $\delta{\cal H}$, the~term in $\cal H$ which
breaks scale invariance explicitly. Type-III theories require
$\delta{\cal H}$ to generate large-scale effects such as
$\Lambda_\textsc{tc}$ and the fermion condensate at the sill, but~to
\emph{desist} in cases where that is inconvenient, e.g.,\ the 
scalar-boson~mass. 

\begin{figure}[H]  
\centering
\includegraphics[width=13.5 cm]{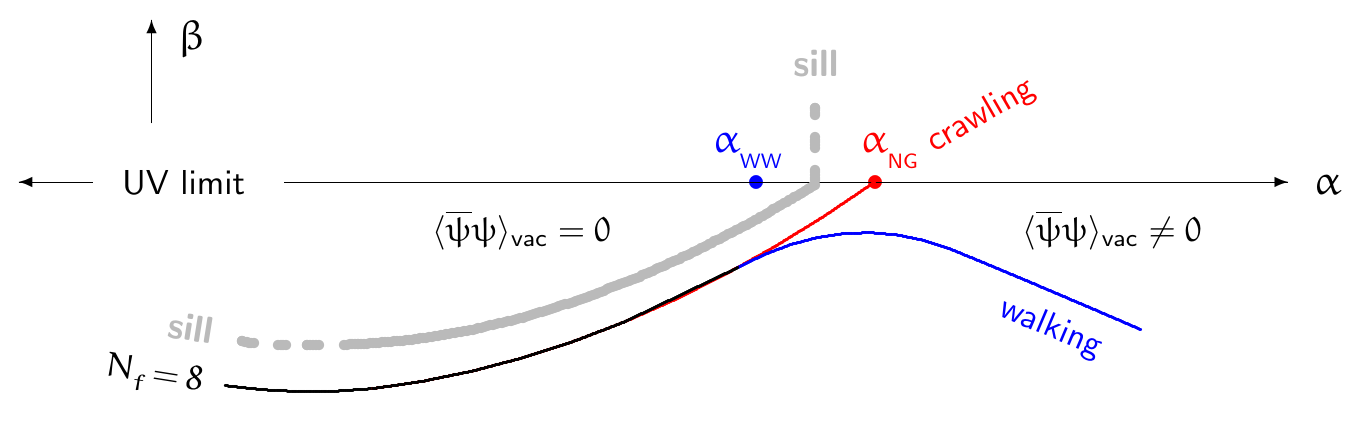}
\caption{Competing explanations for the appearance of
  a scalar particle in lattice data for $SU(3)$ gauge theory with $N_f
  =8$ triplet fermions. Two IRFPs are shown: (a) the closest scale-free 
 IRFP $\alphaWW$ just inside the conformal window, $N_f^\mathrm{c} <
 N_f \leqslant 16$, and~(b) a scale-dependent IRFP $\alphaNG$ for $N_f
 = 8 \leqslant N_f^\mathrm{c}$. Walking TC assumes that $\alphaNG$ is
not present. Instead, the~small scalar mass is supposed to arise at an
intermediate energy where the curve is closest to the axis, and~then the
theory chooses the blue line labelled ``walking'' to approach
the infrared region. In~crawling TC, the~$N_f = 8$ theory enters the
infrared region as it approaches the axis and chooses the red line
labelled ``crawling''. The~short length of the red line accounts for
the small mass acquired by the type-I dilaton.
\label{fig4}}
\end{figure}

The most likely explanation~\cite{crawl} of the light scalar particle
is that there exists an IRFP $\alphaNG$ just outside the conformal
window. Since there is scale dependence at $\alphaNG$,       
a genuine type-I dilaton and hence all large-scale effects exist at that 
point. At~$\alphaNG$, both conformal and chiral invariance are
in the NG mode, so massless technipions exist there as well as the 
type-I dilaton. Large-scale effects cannot be due to $\delta{\cal H}$,
because $\alpha$ can run smoothly to $\alphaNG$ in the conformal limit 
$\delta{\cal H} \to 0$. The~sill does not get in the way, so there is
no need to assume anything about its dynamics. The~small scale of
the scalar-particle mass $M_\sigma$ corresponds to $\alpha$ being in
the infrared region close to $\alphaNG$ (Figure \ref{fig4}). 

In walking TC, IRFPs outside the conformal window are thought to be
forbidden. Instead, an~explanation for the light scalar particle is
sought by appending dilaton Lagrangians to the type-III 
framework~\cite{fodor18,Appel17,Appel18,Appel19,fodor19}.
In fact, these effective Lagrangians are type-I theories developed in 
1970~\mbox{\cite{isham70,ellis70,zumino70}}:
they generate asymptotic expansions in $\delta{\cal H}_\text{eff} \sim
0$ about a conformal limit with scale-dependent amplitudes depending
on the decay constant $f_\sigma$ or its TC analogue
$F_\sigma$. The~question is: if $\alphaNG$ is not available, about
what point is the expansion to be performed?

Since the emphasis in walking TC has been to minimize $|\beta|$,
presumably the understanding has been that the expansion should be
carried out about $\alphaWW$. The~trouble with that is the lack of scale
dependence at $\alphaWW$. An~alternative has just been 
suggested~\cite{golt20}, that the dilaton Lagrangian expansion
should correspond to expanding in ``the distance to the conformal
window'', i.e.,\ about the sill, where there is certainly a large
scale. In~a type-III theory, the sill acquires its scale dependence
from a large-scale violation $\delta{\cal H}$ in the
Hamiltonian. The~problem is then that the expansion of 
${\cal L}_\text{dil}$ is about a point where the 
corresponding effective Hamiltonian is conformally \emph{invariant}.

The conclusion is that type-I and type-III theories should not be
mixed---they are based on contradictory assumptions. A~type-I
effective Lagrangian introduced to discuss the light scalar boson
should be given a type-I point about which it can be expanded:
$\alphaNG$. Why this should be such \mbox{a fearsome} prospect is~puzzling.
 
Finally, I should comment on the perception that lattice data for $N_f
= 4$ implies that the $f_0(500)$ is heavy, contrary to my remarks
below Figure~\ref{spectrum}. A~comparison of data for $N_f = 4$ and
$N_f = 8$ (Figure 1 of~\cite{appel19}) shows that as the fermion mass
$m_\psi$ becomes small,  the~light scalar particle is almost
degenerate with (techni-)pions for $N_f = 8$ but not for $N_f = 4$.
This indicates that the gluonic contribution to the scalar mass is
negligible for $N_f = 8$ but not for $N_f = 4$. A~type-III
interpretation of this is that the gluonic contribution is a
large-scale effect due to $\delta{\cal H}$---a point of view
similar to that of~\cite{bardeen86,hold87,hold88}.

In type-I theories, there is no problem. A~gluonic contribution
to the scalar mass is a small-scale effect due to the coupling
being close to but not at the NG-mode IRFP. For~$N_f = 8$, apparently 
$\epsilon = \alphaNG - \alpha$ is so small that the scalar mass
is dominated by the masses $m_\psi$ used in the lattice analysis.
For $N_f = 4$ (a lattice-friendly approximation to the physical case
of $N_f = 3$ light flavors), the~effects of $\epsilon$ appear similar
in magnitude to those of $m_\psi$, within~fairly large~errors. 

So, I maintain that the Higgs boson is the direct TC analogue of $f_0(500)$ 
for QCD: both are derived from type-I dilatons at
scale-dependent IRFPs. The~main difference is in the ratio $r$ of
small-scale to large-scale effects,
\begin{equation} 
  r_\textsc{qcd} \approx \frac{500\ \text{MeV}}{4\pi(92\ \text{MeV)}} = 0.4 
\quad\mbox{and}\quad 
 r_\textsc{tc} \approx \frac{125\ \text{GeV}}{4\pi(246\ \text{GeV})} =
 0.04 \,.
\end{equation}
This is permissible in type-I theories because large- and small-scale
effects have separate~origins. 
  
\vspace{6pt}
 
\funding{This research received no external~funding.}

\acknowledgments{I thank Lewis Tunstall and Oscar Cat\`{a} for their
  continuing advice after a long and \mbox{fruitful collaboration}.}

\conflictsofinterest{The author declares no conflict of~interest.}

\abbreviations{The following abbreviations are used in this
  manuscript:}

\begin{multicols}{2}
\noindent
\begin{tabular}{@{}ll}
IRFP & infrared fixed point \\
QCD & Quantum Chromodynamics\\
$\chi$PT$_\sigma$  & chiral-scale perturbation theory \\
TC & Technicolor  \\
NG & Nambu–Goldstone \\
WW & Wigner–Weyl\\
$\chi$PT$_3$ & chiral $SU(3)_L \times SU(3)_R$ perturbation theory
\end{tabular}

\noindent
\begin{tabular}{@{}ll}
$\chi$PT$_2$ & chiral $SU(2)_L \times SU(2)_R$ perturbation theory \\
U$\chi$PT & unitarized chiral perturbation theory \\
RG & renormalization group \\
LO & leading order \\
GW & Gildener–Weinberg  \\
CW & Coleman–Weinberg \\
UV & ultraviolet 
\end{tabular}
\end{multicols}


\reftitle{References}


\end{document}